\documentclass[submission,copyright,creativecommons]{eptcs}
\providecommand{\event}{SCSS 2012} 
\usepackage{breakurl}             
\usepackage{amssymb}
\usepackage{epsfig}
\usepackage{multirow}
\usepackage{amsmath}
\usepackage{dsfont}
\usepackage{cite}      
\usepackage{graphicx}  
\usepackage{psfrag}    
\usepackage{subfigure} 
\usepackage{url}       
\usepackage{stfloats}  
\usepackage{array}
\usepackage{multirow}
\usepackage{amssymb}
\usepackage{alltt}
\usepackage[english]{babel}

\title{Formal Analysis of Soft Errors using Theorem Proving}
\author{Naeem Abbasi, Osman Hasan and Sofi\`{e}ne Tahar
\institute{ECE Department, Concordia University, Montreal, QC, Canada\\}
\email{\{n\_ab,o\_hasan,tahar\}@ece.concordia.ca}
}
\def\titlerunning{Formal Analysis of Soft Errors using Theorem Proving}
\def\authorrunning{N. Abbasi, O. Hasan \& S. Tahar}
\begin{document}
\maketitle

\begin{abstract}
Modeling and analysis of soft errors in electronic circuits
has traditionally been done using computer simulations. Computer
simulations cannot guarantee correctness
of analysis because they utilize approximate real number
representations and pseudo random numbers in the analysis and thus are not well suited for analyzing safety-critical applications. In this
paper, we present a higher-order logic theorem
proving based method for modeling and analysis of soft errors in electronic circuits.
Our developed infrastructure includes formalized continuous random
variable pairs, their Cumulative Distribution Function (CDF) properties and independent standard uniform
and Gaussian random variables. We illustrate the usefulness of
our approach by modeling and analyzing soft errors in commonly used
dynamic random access memory sense amplifier circuits.
\end{abstract}

\section{Introduction}
\label{intro_sec}
In many safety critical applications, such as in avionics, electronic equipment operates in harsh environments and experiences extreme temperatures and excessive doses of solar and cosmic radiations. This can often result in change in the state of the charge storage nodes in electronic circuits. Such abnormal changes in the states of storage nodes in electronic circuits are called soft errors~\cite{MainSoftErrorModelLaymanChamberlain}. These nonrecurrent and non permanent errors can cause an electronic system to behave in an un predictable way. There are four commonly known causes of soft errors in logic and memory circuits: 1) undesirable capacitive coupling of circuit elements~\cite{Ref4:RWKeyes}, 2) circuit parameter fluctuations and variations, 3) ionizing particle and EM radiation, and 4) built-in thermal, shot and 1/$f$ noise. Good circuit design and layout techniques can be used to effectively eliminate soft errors due to undesirable capacitive coupling and circuit parameter variations~\cite{Ref7:Masudaet.al.}. In order to deal with the other two types of soft errors accurate analysis of the design is required~\cite{Ref8:MayWoods,MainSoftErrorModelLaymanChamberlain}.

Soft error occurrence mechanism is random in nature and is usually analyzed using simulation based techniques such as Monte carlo simulation methods~\cite{MONTE_CARLO_SER_TOOLS_5_1949}. These techniques tend to be inaccurate and slow and are unsatisfactory for safety critical applications. Realistic analysis of most practical linear and non-linear circuits involves real and random variables. Formal methods based techniques, such as probabilistic model checking, are unsuitable for the analysis of such problems as it is usually not possible to accurately model the continuous electronic circuit behavior using finite state systems.

 In this paper, we apply the higher-order logic theorem proving method~\cite{gordon_89} to the problem of random effect modeling and analysis in electronic circuits. The main reason for using higher-order logic is to leverage upon its high expressiveness, which allows us to precisely model any system that can be expressed mathematically. Thus, it allows us to construct true continuous and randomized models of electronic circuits and thus alleviates the limitations of simulation and model checking based analysis techniques. These models are then used to form an equivalence or an implication relation with their specifications. These boolean relationships are then proved using mathematical reasoning in the sound core of the HOL theorem prover.

Probabilistic analysis infrastructure has been developed in HOL during the last decade. Hurd formalized discrete random variables having Uniform, Bernoulli, Binomial, and Geometric probability mass functions in the HOL theorem prover \cite{hurd_02}. Audebaud et al., describe a method for proving properties of randomized algorithms in Coq proof assistant \cite{pauline09}. They use functional and algebraic properties of unit interval to show the validity of general rules for estimating the probabilities of randomized algorithms. However, similar to Hurd's work, their approach can only address discrete distributions. Hasan, building on Hurd’s work, formalized statistical properties of single and multiple discrete random variables, continuous random variables with various distributions using inverse transform method \cite{hasan_phd_08} and verified their probabilistic and some statistical properties \cite{FM_PAPER_09}. Harrison \cite{HarrisonJ_1_2005} formalized the guage integration on finite-dimensional Euclidean spaces, which is quite similar to product space of Lebesgue measures. Okazaki and Shidama \cite{OkazakiS_ShidamaY_2009} formalized properties of real valued random vairables in Mizar. More recently, Hoelzl \cite{johannes_tphol_11} and Mhamdi \cite{mhamdi_tphol_2010,mhamdi_tphol_11} formalized basic notions of measure, topology and lebesgue integration. These formalisms are based on extended real numbers and are thus more expressive than Hurds formalization of probability theory. However, they do not contain a specific probability space due to which they cannot be used to verify random variable functions. Since, soft error is primarily based on modeling the uncertainties by appropriate random variables so we have chosen Hurd's formalization of measure theory for this work. To the best of our knowledge, the foremost foundations of soft error analysis of electronic circuits, such as the formalization of continuous random variable pair, its classic Cumulative Distribution Function (CDF) properties, and the formalization of Gaussian random variable pair do not exist in literature and is presented for the very first time in this paper.

\begin{figure}[htb!]
  \centering
    \includegraphics[width=0.75\textwidth]{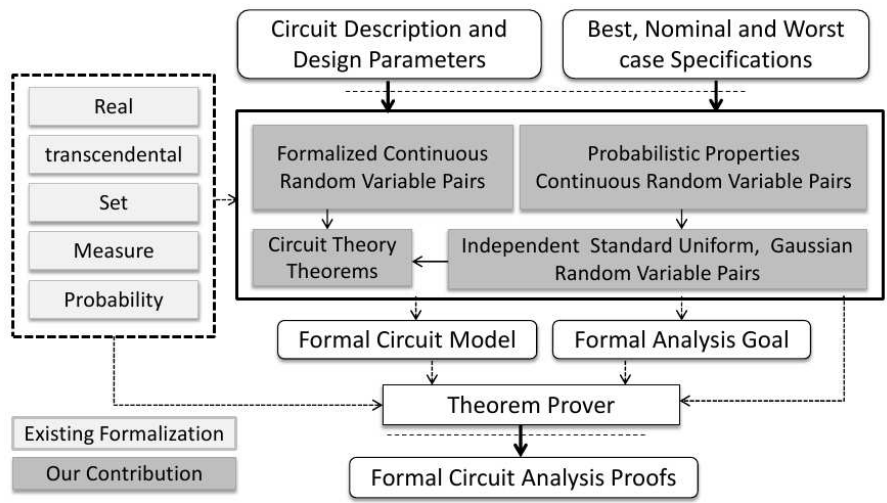}
    \caption{Proposed methodology for circuit analysis}
    \label{fig_method_label}
    \end{figure}

Our proposed method is shown in Figure \ref{fig_method_label}. We build on existing real number, transcendental function, set, measure, and probability theories in the HOL theorem prover. Our developed infrastructure includes formalization of a continuous random variable pair using an approach similar to~\cite{hasan_phd_08}. We have formalized important notions of joint and marginal cumulative distribution functions and the independence of random variable pairs. Using the specification of random variable pairs, we then verify their CDF properties by interactively constructing the proofs of these properties for arbitrary continuous random variables. Then using Inverse Transform Method, we have formalized random variable pairs for which inverse CDF function of the cumulative probability distribution function exists.

In this paper, we also describe the formalization of a pair of independent Gaussian random variables using the Box-Muller method \cite{BoxMuller}. We then utilize these variables in the modeling and formal analysis of soft errors caused by thermal noise in sense amplifier circuit of a DRAM. In a typical analysis using our proposed method, the design and the best, nominal and worst case specifications are first expressed using higher-order logic. Uncertain design and operating environment behaviors are then accurately modeled using formalized random variables in higher-order logic. Design uncertainties include noise and device model parameter variations. Realistic and accurate operating environment uncertainties include effects such as variations in the operating temperature, supply voltage, and varying doses of incident particle and electromagnetic radiation. Finally, the analysis is carried out interactively in the trusted kernel of the HOL theorem prover and formal circuit and system analysis proofs are constructed.

The rest of the paper is organized as follows: Section~\ref{HOL_Formalization} describes the formalization of continuous random variable pair, verification of its classical properties, and formalization of standard Uniform and Gaussian random variable pairs. Using the developed infrastructure, we describe an accurate analysis of soft errors in the sense amplifier of dynamic random access memories in Section~\ref{ApplicationSEC}. Finally, Section~\ref{conclusionSEC} concludes the paper.

\section{Formalization of Continuous Random Variables}
\label{HOL_Formalization}
In Hurd's formalization \cite{hurd_02}, a random variable $\mathcal{F}$ is a higher-order logic probabilistic function which takes a parameter of type $\alpha$ and an infinite Boolean sequence, ranges over values of type $\beta$ and upon termination returns the remaining portion of the infinite Boolean sequence.

\[\mathcal{F}: \alpha \rightarrow B^{\infty} \rightarrow
\beta \times B^{\infty}\]

Hurd formalized four probabilistic algorithms with Uniform, Bernoulli, Binomial, and Geometric probability mass functions. Hasan \cite{hasan_phd_08}, building on Hurd's work, formalized a standard uniform random variable as a special case of the discrete version of a uniform random variable, as given in Equation~\ref{Equation_cont_rv_2}.
\begin{equation}
\label{Equation_cont_rv_2}
\lim_{n \rightarrow \infty}(\lambda n. \sum_{k=0}^{n-1} (\frac{1}{2})^{k+1} X_{k})
\end{equation}
\noindent where $\displaystyle (\lambda n. \sum_{k=0}^{n-1} (\frac{1}{2})^{k+1} X_{k})$ represents the discrete uniform random variable. Hasan's  formal specification of the standard uniform random variable in HOL is given in Definition 1, and is based on Equation~\ref{Equation_cont_rv_2}.
\begin{flushleft}
\texttt{\bf{Definition 1:}} \emph{Standard Uniform Random Variable} \cite{hasan_phd_08}\\
\noindent $\vdash$ $\forall $ \texttt{s. std\_unif\_cont s = $\lim$ ($\lambda$n. fst (std\_unif\_disc n s))}
\end{flushleft}
\noindent The function \texttt{std\_unif\_disc} is a standard discrete uniform random variable in HOL. It takes two arguments, a natural number (\texttt{n}:num) and an infinite sequence of random bits (\texttt{s}:num$\rightarrow$bool). The function utilizes these two arguments and returns a pair of type (real, num$\rightarrow$bool). The real value corresponds to the value of the random variable and the second element in the pair is the unused portion of the infinite boolean sequence. The function \texttt{fst} takes a pair as input and returns the first element of the pair, and the function \texttt{lim P} in HOL is the formalization of the limit of a real sequence \texttt{P}. Using Inverse Transform Method (ITM), Hasan formalized uniform, triangular, exponential and rayleigh random variables. Hasan's HOL formalization of a standard uniform random variable \texttt{uniform\_rv} is as follows:

\texttt{$\vdash \forall \mbox{s a b. uniform\_rv a b s = (b - a)(std\_unif\_cont\ s) + a} $}.

\noindent Where \texttt{a} and \texttt{b} are the two real parameters of the uniform random variable. We build upon these foundations to formalize pairs of continuous random variables in this paper.

\subsection{Continuous Random Variable Pairs}
\label{subsec_form_pairs_crv}
Multiple independent random variables are often required to model and analyze hard to predict and random behaviour of electronic circuits and systems. To perform such modeling and analysis in a theorem proving environment formalized independent random variables are needed. Building on Hasan's work, we formalize a pair of Uniform continuous random variables as:

$\displaystyle (
\lim_{n \rightarrow \infty}(\lambda n. \sum_{k=0}^{n-1} (\frac{1}{2})^{k+1} X_{1k}),
\lim_{n \rightarrow \infty}(\lambda n. \sum_{k=0}^{n-1} (\frac{1}{2})^{k+1} X_{2k})
)$,

\noindent where $\displaystyle (\lambda n. \sum_{k=0}^{n-1} (\frac{1}{2})^{k+1} X_{ik})$, $i \in \{1,2\}$, represents a discrete uniform random variable. The HOL formalization of a pair of Uniform continuous random variables is given by:

\texttt{$\vdash \forall \mbox{s.~std\_unif\_pair\_cont s} = (\displaystyle \lim_{n \rightarrow \infty} (\lambda \mbox{n.~fst (std\_unif\_disc~n~(seven s)))},$}

\texttt{$\displaystyle ~~~~~~~~~~~~~~~~~~~~~~~~~~~~\lim_{n \rightarrow \infty} (\lambda \mbox{n.~fst (std\_unif\_disc n (sodd  ~s)))})$ }

\texttt{$
\vdash \forall \mbox{s. X1\_S\_UNIF s = fst (std\_unif\_pair\_cont s)};$}

\texttt{$
\vdash \forall \mbox{s. X2\_S\_UNIF s = snd (std\_unif\_pair\_cont s)}
$}

We also formalize important concepts of Joint and Marginal Cumulative Distribution Functions and the Independence of a pair of random variables. These concepts play a vital role in analyzing soft errors as will be demonstrated later. Our formalization of these concepts is based on \cite{khazanie_76}.

\noindent Definition 2 describes the HOL formalization of the joint CDF of a pair of random variables mathematically expressed as:

$F_{X_{1},X_{2}}(x_{1},x_{2})=P(X_{1} \leq x_{1} \wedge X_{2} \leq x_{2})$.

\begin{flushleft}
\texttt{\bf{Definition 2:}} \emph{Joint CDF of a Pair of Random Variables}\\
\noindent $\vdash$ $\forall$ \texttt{X1 X2 x1 x2. joint\_cdf X1 X2 x1 x2 = \\
\indent \indent \indent \indent ~~~~~~~~~~~~~~~~~~~~prob bern \{s $|$ (X1 s $\leq$ x1) $\wedge$ (X2 s $\leq$ x2)\}}
\end{flushleft}

\noindent where \texttt{X1} and \texttt{X2} are the first and second element of the random variable pair and \texttt{x1} and \texttt{x2} are two real numbers.

The marginal CDF functions of a pair of random variables \texttt{($X_{1}$,$X_{2}$)} is defined as:

~\noindent \texttt{\mbox{$\displaystyle F_{X1}(x1) = \lim_{x2 \rightarrow \infty}F_{X1,X2}(x1,x2)$} =} \texttt{\mbox{P(X1 $\leq$ x1)}}
and

~\noindent \texttt{\mbox{$\displaystyle F_{X2}(x2) = \lim_{x1 \rightarrow \infty} F_{X1,X2}(x1,x2)$} =}
\texttt{\mbox{P(X2 $\leq$ x2)}}.

\noindent The HOL formalization of marginal CDF functions is given in Definition 3.

\begin{flushleft}
\texttt{\bf{Definition 3:}} \emph{Joint CDF of a Pair of Random Variables}\\
\noindent $\vdash$ $\forall$ \texttt{X1 X2 x1. marginal\_cdf\_x1 X1 X2 x1 =}\\
~~~~~~~~~~~~~~~~~~~~~\texttt{lim (\mbox{$\lambda$}n. prob bern \{s| (X1 s) \mbox{$\leq$} x1 \mbox{$\wedge$} (X2 s) \mbox{$\leq$} (\&n))\})} \\

\noindent $\vdash$ $\forall$ \texttt{X1 X2 x2. marginal\_cdf\_x2 X1 X2 x2 = }\\
~~~~~~~~~~~~~~~~~~~~~\texttt{lim (\mbox{$\lambda$}n. prob bern \{s| (X1 s) \mbox{$\leq$} (\&n) \mbox{$\wedge$} (X2 s) \mbox{$\leq$} x2)\})}
\end{flushleft}


Two random variables \texttt{X1} and \texttt{X2} are said to be independent if for every pair of real numbers \texttt{x1} and \texttt{x2} the two events \{X1 $\leq$ x1\} and \{X2 $\leq$ x2\} are independent. Which means that the value of one random variable has no influence on the other and vice versa. This notion is very useful in accurate and realistic modeling of practical electronic circuits and systems. Mathematically the notion of independence is defined as:

\noindent \texttt{P\{X1 $\leq$ x1 $\wedge$ X2 $\leq$ x2\} = P\{X1 $\leq$ x1\}.P\{X2 $\leq$ x2\}}

The HOL formalization is given in Definition 4.
\begin{flushleft}
\texttt{\bf{Definition 4:}} \emph{Independent Random Variable Pair}\\
\noindent $\vdash$ $\forall $ \texttt{X1 X2 x1 x2. independent\_rv\_pair X1 X2 x1 x2 = \\
~~~~~(\{s | X1 s \mbox{$\leq$} x1 \mbox{$\wedge$} X2 s \mbox{$\leq$} x2\} IN events bern) $\wedge$\\
~~~~~(prob bern \{s | X1 s $\leq$ x1 $\wedge$ X2 s $\leq$ x2\} = \\
~~~~~~prob bern \{s | X1 s $\leq$ x1\} * prob bern \{s | X2 s $\leq$ x2\})}
\end{flushleft}

\subsection{Formal Verification of CDF Properties of Pairs of Random Variables}
\label{subsec_form_verif_cdf_pair_crv}
Using the formal specification of the CDF function for a pair of random variables, we have formally verified the classical properties of the CDF of a pair of random variables. These properties are verified under the assumption that the set \texttt{\{s $|$ R s  x\}}, where \texttt{R} represents a pair of random variables under consideration, is measurable for all values of the pair. The formal proofs for these properties confirm our formalized specifications of the CDF of a pair of random variables. We summarize these results in Table~\ref{rv_table_X}.

\begin{table*}[htb]
\caption{CDF properties of continuous random variable pairs}
\label{rv_table_X}
\centering
\begin{tabular}{|l|l|}  \hline
{\footnotesize{\bfseries Property}} & {\footnotesize{\bfseries Mathematical Description}}; {\footnotesize{\bfseries HOL Formalization}}
\\ \hline
\footnotesize{
$\begin{array}{ll}
\texttt{CDF} \\
\texttt{Bounds}
\end{array}$}
&
\footnotesize{
$\begin{array}{ll}
\noindent \texttt{$0 \leq F_{X_{1},X_{2}}(x_{1},x_{2}) \leq 1$;
}\\
\noindent \vdash \forall\texttt{X1 X2 x1 x2.}
\texttt{CDF\_pair\_in\_events\_bern X1 X2 x1 x2 \mbox{$\Rightarrow$}} \\
\texttt{((0 \mbox{$\leq$} joint\_cdf X1 X2 x1 x2) \mbox{$\wedge$} (joint\_cdf X1 X2 x1 x2 \mbox{$\leq$} 1))}
\end{array}$}
\\ \hline
\footnotesize{
$\begin{array}{ll}
\texttt{CDF}\\
\texttt{Monotonic} \\
\texttt{Non}\\
\texttt{decreasing}
\end{array}$}
&
\footnotesize{
$\begin{array}{ll}
\noindent \texttt{\mbox{$F_{X_{1},X_{2}}(a,c)$}}
~\texttt{\mbox{$\leq$}} \texttt{\mbox{$F_{X_{1},X_{2}}(b,d)$}};\\
\noindent \vdash \forall \texttt{a b c d.~(a \mbox{$<$} b) \mbox{$\wedge$} (c \mbox{$<$} d) \mbox{$\wedge$}} \\
\texttt{(\mbox{$\forall$} x1 x2. CDF\_pair\_in\_events\_bern X1 X2 x1 x2) \mbox{$\Rightarrow$} }\\
\texttt{( (joint\_cdf X1 X2 a c \mbox{$\leq$} joint\_cdf X1 X2 b c) \mbox{$\wedge$}} \\
~~~\texttt{(joint\_cdf X1 X2 b c \mbox{$\leq$} joint\_cdf X1 X2 b d) ) }

\end{array}$}
\\ \hline
\footnotesize{
$\begin{array}{ll}
\texttt{CDF Pair} \\
\texttt{at +$\infty$}
\end{array}$}
&
\footnotesize{
$\begin{array}{ll}
\noindent \texttt{\mbox{$\displaystyle\lim_{x2\rightarrow\infty}$} \mbox{$\displaystyle  \lim_{x1\rightarrow\infty}$}} \texttt{\mbox{$F_{X1,X2}$}(x1,x2) =}
\texttt{$F_{X1,X2}$($\infty$,$\infty$) = 1};\\
\noindent \vdash \texttt{ (\mbox{$\forall$} X1 x1. CDF\_in\_events\_bern X1 x1) \mbox{$\wedge$}} \\
\texttt{(\mbox{$\forall$} X1 X2 x1 x2. CDF\_pair\_in\_events\_bern X1 X2 x1 x2) \mbox{$\Rightarrow$}} \\
\texttt{(lim (\mbox{$\lambda$}n1. lim (\mbox{$\lambda$}n2. joint\_cdf X1 X2 (\& n1) (\& n2))) = 1)}
\end{array}$}
\\ \hline
\footnotesize{
$\begin{array}{ll}
\texttt{CDF Pair} \\
\texttt{at -$\infty$}
\end{array}$}
&
\footnotesize{
$\begin{array}{ll}
\noindent \texttt{$\displaystyle \lim_{x2 \rightarrow -\infty} F_{X1, X2}(x1,x2)$ =} \noindent \texttt{$\displaystyle \lim_{x1 \rightarrow -\infty} F_{X1, X2}(x1, x2)$ = 0};\\
\noindent \vdash \texttt{ (\mbox{$\forall$}X1 X2 x1 x2. CDF\_pair\_in\_events\_bern X1 X2 x1 x2) \mbox{$\Rightarrow$}} \\
\texttt{( (lim (\mbox{$\lambda$}n. joint\_cdf X1 X2 (- \& n) x2) = 0) \mbox{$\wedge$}} \\
\texttt{(lim (\mbox{$\lambda$}n. joint\_cdf X1 X2 x1 (- \& n)) = 0) )}
\end{array}$}
\\ \hline
\end{tabular}
\end{table*}

As an example, we present the proof of one such property here called the CDF interval property. The rest of the formal proofs can be found in \cite{abbasi_10_pair_rv_verification_tr}.

\noindent \textbf{CDF Pair Interval Property}

\noindent If \texttt{a}, \texttt{b}, \texttt{c}, and \texttt{d} are real numbers with \texttt{a \mbox{$<$} b}, and \texttt{c \mbox{$<$} d}, then the probability of an interval event of a pair of random variables is given by \texttt{P(a $<$ X1 $\leq$ b, c $<$ X2 $\leq$ d) = \mbox{$F_{X1,X2}$}(b,d) - \mbox{$F_{X1,X2}$}(b,c) - \mbox{$F_{X1,X2}$}(a,d) + \mbox{$F_{X1,X2}$}(a,c)}. The property is formally stated in Theorem 1.

\begin{flushleft}
\texttt{\bf{Theorem 1:}} \emph{CDF Pair Useful Interval Property}\\
\noindent $\vdash$ \texttt{$\forall$a b c d. (a $<$ b) $\wedge$ (c $<$ d) $\wedge$ \\
\indent \indent ~\{s | X1 s $\leq$ a $\wedge$ c $<$ X2 s $\wedge$ X2 s $\leq$ d\} IN events bern $\wedge$ \\
\indent \indent ~\{s | a $<$ X1 s  $\wedge$ X1 s $\leq$ b $\wedge$ c $<$ X2 s $\wedge$ X2 s $\leq$ d\} IN events bern \\
\indent \indent ~\{s | X1 s $\leq$ a $\wedge$ X2 s $\leq$ c\} IN events bern $\wedge$ \\
\indent \indent ~\{s | X1 s $\leq$ b $\wedge$ X2 s $\leq$ c\} IN events bern $\wedge$ \\
\indent \indent ~\{s | X1 s $\leq$ b $\wedge$ c $<$ X2 s $\wedge$ X2 s $\leq$ d\} IN events bern $\Rightarrow$ \\
\indent \indent ~~( prob bern \{s | a $<$ X1 s $\wedge$ X1 s $\leq$ b $\wedge$ c $<$ X2 s $\wedge$ X2 s $\leq$ d\} = \\
\indent \indent ~~~~joint\_cdf X1 X2 b d - joint\_cdf X1 X2 b c - \\
\indent \indent ~~~~joint\_cdf X1 X2 a d + joint\_cdf X1 X2 a c ) }
\end{flushleft}

\textbf{Proof:} The proof of this property begins by first showing that the events \texttt{(a \mbox{$<$} X1 \mbox{$\leq$} b \mbox{$\wedge$} c \mbox{$<$} X2 \mbox{$\leq$} d)} and \texttt{(X1 \mbox{$\leq$} a \mbox{$\wedge$} c \mbox{$<$} X2 \mbox{$\leq$} d)} are disjoint. Then we show that
P(a \mbox{$<$} X1 \mbox{$\leq$} b \mbox{$\wedge$} c \mbox{$<$} X2 \mbox{$\leq$} d) +
P(X1 \mbox{$\leq$} a \mbox{$\wedge$} c \mbox{$<$} X2 \mbox{$\leq$} d)
=
P(X1 \mbox{$\leq$} b \mbox{$\wedge$} c \mbox{$<$} X2 \mbox{$\leq$} d), using the additive law of probabilities \cite{hurd_02}. Similarly, we prove that,
P(X1 \mbox{$\leq$} b \mbox{$\wedge$} c \mbox{$<$} X2 \mbox{$\leq$} d) +
P(X1 \mbox{$\leq$} b \mbox{$\wedge$} X2 \mbox{$\leq$} c)
=
P(X1 \mbox{$\leq$} b \mbox{$\wedge$} X2 \mbox{$\leq$} d)
and
P(X1 \mbox{$\leq$} a \mbox{$\wedge$} c \mbox{$<$} X2 \mbox{$\leq$} d) +
P(X1 \mbox{$\leq$} a \mbox{$\wedge$} X2 \mbox{$\leq$} c)
=
P(X1 \mbox{$\leq$} a \mbox{$\wedge$} X2 \mbox{$\leq$} d). Finally, we conclude the proof by rewriting and simplifying with the definitions of the joint CDF function and the above results. This property states that the probability that the random vector \texttt{(X1,X2)} falls in a rectangular region and can be found by combining the values of cumulative distribution function at the four corners of the rectangular region.

\subsection{Formalization of Gaussian Random Variable Pairs}
\label{subsec_form_pairs_crv}

Thermal noise in electronic circuits is caused by random motion of electrons in semiconductor materials and is typically modeled using Gaussian random variables. In this section, we describe the HOL formalization of a pair of independent Gaussian random variables using the Box-Muller method \cite{BoxMuller}. According to the Box-Muller method, given a pair of independent standard Uniform random variables \texttt{($U_{1}$,$U_{2}$)}, a pair of independent Gaussian random variable can be formalized as:

\texttt{($G_{1}$,$G_{2}$) = ($\sqrt{-2~ln~U_{1}}~cos(2~\pi~U_{2})$,$\sqrt{-2~ln~U_{1}}~sin(2~\pi~U_{2})$)}.

\noindent The HOL formalization of the Gaussian random variable is given in Table~\ref{gausssian_rv_table}.

\begin{table*}[htb]
\caption{\footnotesize{Gaussian random variable formalization in HOL}}
\label{gausssian_rv_table}
\centering
\begin{tabular}{|l|l|}  \hline
{\footnotesize{\bfseries Distribution}} & {\footnotesize{\bfseries Formalized Random Variable Pair}}\\ \hline
\footnotesize{
$\begin{array}{ll}
\texttt{Standard} \\
\texttt{Gaussian}\\
\texttt{(0,1)}
\end{array}$}
&
\footnotesize{
$\begin{array}{ll}
\noindent \vdash \forall \texttt{s.~std\_g\_pair\_rv s =} \\
\noindent \texttt{((\mbox{$\sqrt{\mbox{-2~ln~(X1\_S\_UNIF s)}}$} \mbox{$\cos$}(2\mbox{$\pi$}(X2\_S\_UNIF s))),}\\
\noindent \texttt{~(\mbox{$\sqrt{\mbox{-2~ln~(X1\_S\_UNIF s)}}$} \mbox{$\sin$}(2\mbox{$\pi$}(X2\_S\_UNIF s))))}
\end{array}$}  \\ \hline
\footnotesize{
$\begin{array}{ll}
\texttt{Gaussian} \\
\texttt{($\sigma$,$\mu$)}
\end{array}$}
&
\footnotesize{
$\begin{array}{ll}
\noindent \vdash \forall \texttt{\mbox{s $\mu~\sigma$.~g\_pair\_rv $\mu~\sigma$ s}} =\\ \texttt{(\mbox{$\mu$ + $\sigma$~fst (std\_g\_pair\_rv s)}, \mbox{$\mu$ + $\sigma$~snd (std\_g\_pair\_rv s)})}\\
\noindent \vdash \forall \texttt{\mbox{s. V1\_G $\mu$ $\sigma$ s = fst (g\_pair\_rv $\mu$ $\sigma$ s)}};\\
\noindent \vdash \forall \texttt{\mbox{s. V2\_G $\mu$ $\sigma$ s = snd (g\_pair\_rv $\mu$ $\sigma$ s)}}
\end{array}$}  \\ \hline
\end{tabular}
\end{table*}

In this section, we described the formalization of a standard uniform random variable pair and the formalization of a Gaussian random variable pair. These two distributions are used in modeling of realistic process, supply voltage and temperature variations in electronic circuits. Using the formalization described in this section, we can for the very first time model and analyze behavior of analog and mixed signal circuits in a higher-order logic theorem prover and reason about their functional properties in the presence of random process and environment variations. To illustrate the usefulness of our formalization, we present an application in the next section.

\section{Formal Analysis of Soft Errors in DRAMs}
\label{ApplicationSEC}
\subsection{Dynamic Random Access Memory}
\label{DRAM_SubSec}
\begin{figure}[htb!]
  \centering
  \includegraphics[width=1.0\textwidth]{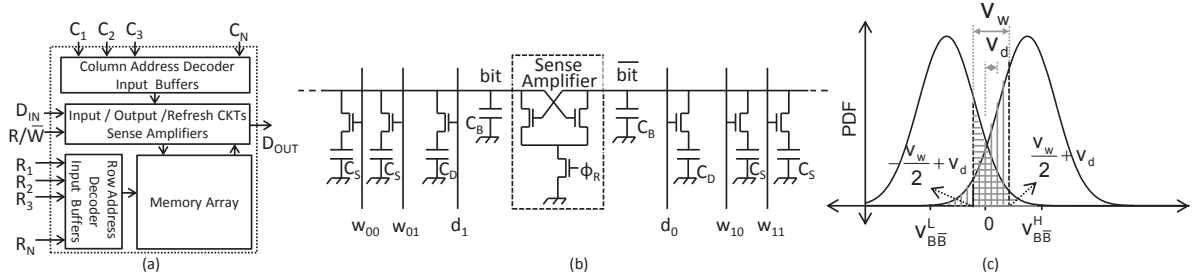}
  \caption{DRAM block diagram (a), balanced bit-line architecture (b), PDF of non-ideal sense amplifier bit line voltages \cite{MainSoftErrorModelLaymanChamberlain} (c).}
    \label{figure2_label}
\end{figure}

Figure~\ref{figure2_label}(a) shows a typical block diagram of a Dynamic Random Access Memory or DRAM. It consists of address buffers, decoders, memory array, and input/output interface circuits. Sense amplifiers are very sensitive differential amplifiers. A differential amplifier usually has three inputs. A pair of inputs is connected to the two bit lines (Figure~\ref{figure2_label}(b), lines bit,$\overline{\mbox{bit}}$). The third input is used to enable the sense amplifier ($\phi_{\mbox{R}}$). The amplifier increases the amplitude of the difference signal between the two bit lines and thermal noise can affect the operation of the sense amplifier. Figure~\ref{figure2_label}(b) shows a balanced bit-line architecture of a commercial DRAM. In this architecture one sense amplifier connects to the bit line of two identical arrays. The circuit diagram shows one transistor storage cells, $C_{S}$, dummy cells, $C_{D}$, and the sense amplifier. The loading effects of the pre-charge, refresh and the input output devices of the DRAM are included in $C_{B}$. More details can be found in \cite{DRAMBook}.

We model the voltages on the two bit lines connected to the inputs of a non-ideal sense amplifier as two independent Gaussian random variables $V_{1}\_G$(-$V_{B\overline{B}}^{L}$,$v_{B\overline{B}n}$) and $V_{2}\_G$($V_{B\overline{B}}^{H}$,$v_{B\overline{B}n}$), where $v_{B\overline{B}n}$ represents the standard deviation of the thermal noise \cite{MainSoftErrorModelLaymanChamberlain}. Figure~\ref{figure2_label}(c) shows the probability density functions (PDF) for the two inputs to the sense amplifier. The vertical shaded area represents the probability of detecting a logic ``1" in the DRAM cell due to the noise when in fact a logic ``0" is stored in that location. Similarly, the horizontally shaded region corresponds to detecting a logic ``0" when in fact a logic ``1" is stored in the memory. The probabilities of a low level being detected as high and that of a high level being detected as low, at the two bit lines, is given by,
$P(-\frac{v_{w}}{2}+v_{d} < V_{1}\_G) = Q \left(\frac{-\frac{v_{w}}{2}+v_{d}-(-V_{B\bar{B}}^{L})}{\sqrt{{v_{B\bar{B}}^{2}}}}\right)$, and $P(V_{2}\_G \leq \frac{v_{w}}{2}+v_{d}) = 1 - Q\left( \frac{\frac{v_{w}}{2}+v_{d}-V_{B\bar{B}}^{H}}{\sqrt{{v_{B\bar{B}}^{2}}}}\right)$, respectively. Where the insensitivity width and the sensitivity center deviation are given by $v_{w}=\delta V_{B\bar{B}}$ and $v_{d}=\chi V_{B\bar{B}}$, where $0\leq \chi,\delta \leq 1$ \cite{MainSoftErrorModelLaymanChamberlain}. Using these assumptions and that both 0 and 1 errors are equally likely to occur, the soft error rate is given by:
$P_{error} = \frac{1}{4} \mbox{erfc} \left[ \frac{V_{B\bar{B}}^{L}}{\sqrt{2} \sqrt{{v}_{B\bar{B}}^{2}}} \left( 1 - \frac{\delta}{2}+\chi \right) \right]
+
\frac{1}{4} \mbox{erfc} \left[ \frac{V_{B\bar{B}}^{H}}{\sqrt{2} \sqrt{{v}_{B\bar{B}}^{2}}} \left( 1 - \frac{\delta}{2}-\chi \right) \right]$. Where error function (\emph{erfc}) is defined as: $\displaystyle erfc(x) = 2 Q\left(\sqrt{2} x \right)$. Next, we formally verify this result using the proposed formalization.
\subsection{Verification of soft error rates}
\label{VerificationSE_SubSec}
\begin{flushleft}
Based on the proposed methodology described in Section 1, the first step is to formally represent the Non-ideal sense amplifier soft error rate model, which can be done as follows:

\texttt{\bf{Definition 4:}} \emph{Non-ideal Sense Amplifier SER Model}\\
\vspace{1pt} \texttt{$\vdash$ $\forall$
$V_{B\bar{B}}^{L}$ $V_{B\bar{B}}^{H}$ $v_{B\bar{B}n}$~$v_{w}$ $v_{d}$.\\
non\_ideal\_ser~$V_{B\bar{B}}^{L}$~$V_{B\bar{B}}^{H}$~$v_{B\bar{B}n}$~$v_{w}$ $v_{d}$ =\\ $\frac{1}{2}$$(\mathbb{P}\{s|(v_{d}-\frac{v_{w}}{2})<(V_{1}\_G~(-V_{B\bar{B}}^{L})~v_{B\bar{B}n}~s)\}$~+ \\
$\mathbb{P}\{s|(V_{2}\_G~V_{B\bar{B}}^{H}~v_{B\bar{B}n}~s)\leq(v_{d}+\frac{v_{w}}{2})\})$}
\end{flushleft}

Now based on this formal definition, we can formally verify the following useful probabilistic relationship regarding the soft error rate for a non-ideal sense amplifier in the presence of thermal noise and parameter variations.

\begin{flushleft}
\texttt{\bf{Theorem 2:}} \emph{Non-ideal Sense Amplifier Soft Error Rate}\\
\vspace{1pt} \texttt{$\vdash$ $\forall$ a b f $V_{B\bar{B}}^{H}$ $V_{B\bar{B}}^{L}$ $v_{B\bar{B}n}$ $\delta$ $\chi$. \\
           ((a$\leq$b) $\wedge$ ($\forall$x. (a$\leq$x) $\wedge$ (x$\leq$b) $\Rightarrow$ (f diffl ($\lambda$t. $\frac{1}{\sqrt{2 \pi}}e^{-\frac{t^{2}}{2}}$) x) x) $\wedge$  (Q1 a b = f b - f a) $\wedge$
           (Q y = $\displaystyle \lim_{n \rightarrow \infty}$ ($\lambda$n. Q1 y (\&n))) $\wedge$ \\
           ($\forall$x. Q x = $\frac{1}{2}$ erfc ($\frac{x}{\sqrt{2}}$)) $\wedge$ 
           ($\forall$z $\mu$ $\sigma$.
              (0 < $\sigma$) $\Rightarrow$
              ($\mathbb{P}$\{s $|$ z < $V_{1}\_G$ $\mu$ $\sigma$ s\} =
               Q ($\frac{z - \mu}{\sigma}$)) $\wedge$
           ($\forall$z $\mu$ $\sigma$.
              (0 < $\sigma$) $\Rightarrow$
              ($\mathbb{P}$\{s $|$ z < $V_{2}\_G$ $\mu$ $\sigma$ s\} =
               Q ($\frac{z - \mu}{\sigma}$)) $\wedge$
              (0$\leq$$\delta$) $\wedge$
           ($\delta$$\leq$1) $\wedge$ (0$\leq$$\chi$) $\wedge$ ($\chi$$\leq$1) $\wedge$ ($v_{w}$ = $\delta$$V_{B\bar{B}}^{H}$) $\wedge$
           ($v_{d}$ = $\chi$ $V_{B\bar{B}}^{H}$) $\wedge$
            (0 < $v_{B\bar{B}n}$) $\wedge$ ($V_{B\bar{B}}^{L} = -V_{B\bar{B}}^{H}$) $\wedge$ (Q(y) + Q(-y) = 1) $\Rightarrow$
           non\_ideal\_ser $V_{B\bar{B}}^{L}$ $V_{B\bar{B}}^{H}$ $v_{B\bar{B}n}$ =
           $\frac{1}{4}$erfc$\left(\frac{V_{B\bar{B}}^{H}}{\sqrt{2}~v_{B\bar{B}n}}\left[ 1 - \frac{\delta}{2} + \chi \right]\right)$ +
            $\frac{1}{4}$erfc$\left(\frac{V_{B\bar{B}}^{L}}{\sqrt{2}~v_{B\bar{B}n}}\left[ 1 - \frac{\delta}{2} - \chi \right]\right)$}
\end{flushleft}

The predicate \texttt{((f diffl ($\lambda$t. $\frac{1}{\sqrt{2 \pi}}e^{-\frac{t^{2}}{2}}$) x) x)} in the first assumption states that the differential of the function \texttt{f} with respect to \texttt{x} is the function ($\lambda$t. $\frac{1}{\sqrt{2 \pi}}e^{-\frac{t^{2}}{2}}$). The second assumption states that \texttt{Q1} is a function with two real arguments \texttt{a} and \texttt{b}, and it returns a real value \texttt{f(b) - f(a)}, which is equal to the value of the definite integral of ($\lambda$t. $\frac{1}{\sqrt{2 \pi}}e^{-\frac{t^{2}}{2}}$). The third assumption then formally represents the \texttt{Q} function as the limit value of function \texttt{Q1} when its second argument tends to infinity. The fourth assumption describes the relationship between the \texttt{Q} function and the error function (\texttt{erfc}, defined in \cite{abbasi_10_pair_rv_verification_tr}). Assumptions 5 and 6 explicitly state that the probabilities of the random variables $V_{1}\_G$ and $V_{2}\_G$ taking values greater than an arbitrary real number \texttt{z} is given by \texttt{Q ($\frac{z - \mu}{\sigma}$)}. Assumptions 7, 8, 9, 10, 11, and 12 state that $\delta$ and $\chi$ which relate the insensitivity width ($v_{w}$ = $\delta$$V_{B\bar{B}}^{H}$) and the sensitivity deviation ($v_{d}$ = $\chi$ $V_{B\bar{B}}^{H}$) parameters to the mean values of the Gaussian random variables $V_{1}\_G$ and $V_{2}\_G$, are real numbers and can only take values in the closed real interval [0,1]. The  thirteenth assumption makes sure that the standard deviation of the thermal noise is a non zero positive value ($0 < v_{B\bar{B}_{n}}$ ). The fourteenth assumption ($V_{B\bar{B}}^{L} = -V_{B\bar{B}}^{H}$) states that the sense amplifier at its inputs sees two equal and opposite polarity dc signals represented by $V_{B\bar{B}}^{H}$ and $V_{B\bar{B}}^{L}$, respectively. The fifteenth assumption states an important property of the Q function that the total area under the Q function is equal to 1.

\textbf{Proof:} We begin the proof by rewriting the right hand side of Theorem 2 with the definition of the complementary error function \texttt{($\forall$x. Q x = $\frac{1}{2}$ erfc ($\frac{x}{\sqrt{2}}$))}, the property of Q function (Q(x)+Q(-x)=1), and three other assumptions of Theorem 2, that is, $v_{w}$ = $\delta$$V_{B\bar{B}}^{H}$, $v_{d}$ = $\chi$ $V_{B\bar{B}}^{H}$ and $V_{B\bar{B}}^{L}=-V_{B\bar{B}}^{H}$. This reduces the righthand side of the proof goal to: \texttt{$\frac{1}{2} \left[1 - \mathbb{P} \{s | (v_{d}+\frac{v_{w}}{2})~<~(V_{2}\_G~V_{B\bar{B}}^{H}~v_{B\bar{B}n}~s)\} \right]$ + $\frac{1}{2} \mathbb{P} \{ s | (v_{d}-\frac{v_{w}}{2}) < (V_{1}\_G~(V_{B\bar{B}}^{H})~v_{B\bar{B}n}~s)\}$}. Now using the fact that $\mathbb{P}(x \leq a)+\mathbb{P}(a < x)=1$, we rewrite the first term in the above expression as:

\noindent \texttt{$\frac{1}{2} \left[\mathbb{P} \{s | (V_{2}\_G~V_{B\bar{B}}^{H}~v_{B\bar{B}n}~s)~\leq~(v_{d}+\frac{v_{w}}{2})\} \right]$ +}\\
\noindent \texttt{$\frac{1}{2} \mathbb{P} \{ s | (v_{d}-\frac{v_{w}}{2}) < (V_{1}\_G~(-V_{B\bar{B}}^{L})~v_{B\bar{B}n}~s)\}$}.

Finally, rewriting the left hand side of the proof goal with the definition of the \texttt{non\_ideal\_ser} and the assumption $V_{B\bar{B}}^{L}=-V_{B\bar{B}}^{H}$, we conclude the proof. More detailed description of the proof can be found in \cite{abbasi_10_pair_rv_verification_tr}.

The HOL code describing our formalization and the soft error rate analysis consists of approximately 1800 lines of code and took over 100 man-hours to complete. The results we presented are guaranteed to be accurate, unlike the simulation based analysis, and are generic due to the universally quantified variables. Such analysis was not possible in the HOL theorem prover earlier because of lack of formalization of pairs of continuous standard uniform and Gaussian random variables which is one of the contributions of this work.
\section{Conclusion}
\label{conclusionSEC}
In this paper, we presented a method for formal analysis of soft errors in electronic circuits using real and independent random variables. We presented the formalization of independent continuous random variable pairs with Uniform and Gaussian distributions. We described soft error rate analysis of a non-ideal sense amplifier circuit commonly used in DRAMs.

Our formalization of Gaussian random variable can be used to perform bit error rate analysis of communication receivers utilizing various modulation schemes such as ASK, PSK and QAM modulations in the presence of additive white Gaussian noise. We are currently working on formalization of lists of independent random variables  to be able to tackle problems with more than two random variables in HOL.

\nocite{*}
\bibliographystyle{eptcs}
\bibliography{scss2012biblio}
\end{document}